\documentclass{eptcs}
\usepackage[utf8]{inputenc}

\usepackage{graphicx}

\usepackage{amsmath}
\usepackage{amssymb}
\usepackage{stmaryrd}
\usepackage{multirow}

\usepackage{bussproofs}
\usepackage{array}
\usepackage {bsymb,b2latexCFP}

\newcommand{\SELECT}{\text{\textbf{SELECT}}\xspace}
\newcommand{\THEN}{\text{\textbf{THEN}}\xspace}

\newcommand{\ENDB}{\text{\textbf{END}}}

\newcommand{\PRE}{\text{\textbf{PRE}}\xspace}

\usepackage{xspace}

\def\ie{{\it i.e.}\xspace}

\def\eg{{\it e.g.}\xspace}

\def\QED{\mbox{\rule[0pt]{1.5ex}{1.5ex}}}
\reversemarginpar

\newcommand{\astd}{\mbox{\textmd{\textsc{astd}}}\xspace}

\newcommand{\eventB}{Event-B\xspace}

\newcounter{ernbcnter}

\newcommand{\autk}{\textsf{\small aut}}

\newcommand{\stdec}{\raisebox{-0.25ex}{$\scriptstyle \circ$}}
\newcommand{\autStk}{\autk\stdec}

\newcommand{\lock}{\textsf{\small \small loc}}

\newcommand{\finalTrans}{\mathit{final}\mbox{?}}

\newcommand{\initk}{\mathit{init}}

\newcommand{\mm}{\hspace*{2em}}

\graphicspath{{img/}}

\title{Formal refinement of extended state machines}

\author{Thomas Fayolle
  \institute{GRIL}
  \email{Thomas.Fayolle@USherbrooke.ca}
  \institute{Université Paris-Est, LACL}
  \email{tfayolle@lacl.fr}
  \institute{Ikos Consulting,\\
    155 rue Anatole France,\\
    92300 Levallois-Perret}
  \email{tfayolle@ikosconsulting.com}
  \and
  Marc Frappier
    \institute{GRIL,\\
    Département Informatique\\
    Université de Sherbrooke,\\
    2500 boulevard université,\\
    Sherbrooke J1K 2R1,\\
    Québec,\\
    Canada }
    \email{Marc.Frappier@USherbrooke.ca}
  \and
  Régine Laleau \qquad  Frédéric Gervais
  \institute{Université Paris-Est, LACL,\\
    IUT Sénart Fontainebleau, Département Informatique,\\
    Route Hurtault,\\
    77300 Fontainebleau, France}
  \email{Frederic.Gervais@u-pec.fr \qquad Laleau@u-pec.fr}
  }

\def\titlerunning{Formal refinement of extended state machines}
\def\authorrunning{Thomas Fayolle, Marc Frappier, Régine Laleau \& Frédéric Gervais}

\begin{document}
\maketitle

\begin{abstract}
  In a traditional formal development process, e.g. using the B method, the informal user requirements are (manually) translated into a global abstract formal specification. This translation is especially difficult to achieve. The Event-B method was developed to incrementally and formally construct such a specification using stepwise refinement. Each increment takes into account new properties and system aspects. In this paper, we propose to couple a graphical notation called Algebraic State-Transition Diagrams (\astd) with an Event-B specification in order to provide a better understanding of the software behaviour. The dynamic behaviour is captured by the \astd, which is based on automata and process algebra operators, while the data model is described by means of an Event-B specification. We propose a methodology to incrementally refine such specification couplings, taking into account new refinement relations and consistency conditions between the control specification and the data specification. We compare the specifications obtained using each approach for readability and proof complexity. The advantages and drawbacks of the traditional approach and of our methodology are discussed. The whole process is illustrated by a railway CBTC-like case study. Our approach is supported by tools for translating \astd's into B and Event-B into B.
\end{abstract}


\section{Introduction}

Specifying a system with formal languages is not straightforward. Our main objective is to specify a whole safety-critical system by using only formal notations and techniques. For validating the whole modelling process, we focus on a railway case study.

The new methodology introduced in this paper is based on the coupling of formal notations and on the joint refinement of both parts of the model. The choice of a specification language is often difficult and depends on the characteristics of the system to be specified. Most often, several languages are good candidates, but none of them,
if taken alone, really fits well, because some aspects of the requirements would not be explicitely taken into account. For instance, when we first consider a state-based formal language like B~\cite{Abrial1996}, safety properties are well captured, but dynamic properties like ordering or liveness properties are not straightforward to express and verify. Likewise, an event-based formal language like CSP~\cite{hoare} 
or LOTOS~\cite{bolognesi} is more convenient for representing dynamic properties, but data models are difficult 
to capture in these languages. In addition to this issue, some syntactical, semantical or technical specificities of a language may constrain a specifier in describing the system. For instance, if there is no modularity or refinement, some systems are difficult to model \emph{ex nihilo}.

In those circumstances, the coupling of specification languages may bring a solution. The aim is to take advantage of the benefits of each notation. However, coupling two distinct formal languages is not an easy task. Syntactically, such an approach must provide a way for reusing the existing notations, especially language operators, in order to be easily understood by people used to specify with one or the other language. Semantically, the different parts of the model must be consistent, so the approach must provide techniques in order to ensure that. Technically, existing tools supporting one or the other language should be reused, as much as possible.

One of the main issues with such couplings comes from the development of the whole system. Concepts like refinement are then required. Instead of writing a complete specification for the whole system in one single step, the different features are specified step by step. In the context of coupling of specification languages, these techniques are even more tricky. For instance, refinement of one part of the model must not introduce inconsistencies with respect to the other part. Sometimes, refinement is considered only in one part of the model in order to prevent that. An embedding of one language into the other one then ensures the consistency of the whole model. Recent work~\cite{oliveira2005refinement,DBLP:conf/asm/IliasovTLRVIL10,DBLP:journals/scp/SchneiderT11,DBLP:journals/fac/Boiten14} try either to define a formal language with a unifying semantics, or to define proof obligations generation rules for showing in which conditions one piece of specification can be refined without introducing inconsistencies.

In our approach, a formal graphical notation, called \astd~\cite{FrGeLaFrSt:2008:revueISSE}, is combined 
with B-like state-based formal specifications for describing the system. The dynamic behaviour is captured 
with an \astd, which is based on automata and process algebra operators, while the data model is described by 
means of B-like specifications. These formal languages will be detailed in Section~\ref{sec:background}.
As a main contribution, we explore complementarity and consistency between \astd and B-like refinements.
Section~\ref{sec:overview} introduces the main principles of our methodology. The case study is 
detailed in Sect.~\ref{sec:casestudy}. Section~\ref{sec:conclusion} concludes the paper with some 
perspectives.    

\section{Background}
\label{sec:background}

\subsection{B and \eventB}
\label{sec:eventb}

B is a formal method~\cite{Abrial1996} supporting the main stages of the software development 
life cycle. Specifications are composed of abstract machines, which encapsulate state variables, 
an invariant constraining the state variables, an initialisation of all the state variables, 
and operations on the state variables. 
The invariant is a first-order predicate in a simplified version of the ZF-set theory, enriched by 
many relational operators. Abstract sets or enumerated sets are used for typing the state variables. 
In B, state variables are modified only by means of substitutions. The initialisation and the operations 
are specified in a generalisation of Dijkstra's guarded command notation, called the Generalised 
Substitution Language (GSL), that allows the definition of non-deterministic and preconditioned 
substitutions. An operation is generally a preconditioned substitution, of the form PRE~$P$~THEN~$S$~END, 
where $P$ is the precondition and $S$ is a substitution. The state transition specified by a preconditioned substitution is guaranteed only when the precondition is satisfied. The main substitutions that will 
be used in the case study are: assignment substitution (denoted by $:=$); substitution of the form 
$x :| (P)$, which states that state variable $x$ is updated such that predicate $P$ becomes true; 
 simultaneous substitutions ($||$); finally, \SELECT substitutions defines many substitutions, each one being guarded by a predicate.

Through refinement steps, the initial abstract machine is transformed, step by step, into a B model 
of the code. Translation
tools are then available for synthesising the final code. Proof activity consists in proving all 
the generated proof obligations for the abstract machine and for each refinement step. In that aim,
the B method is supported by several tools like Atelier B\footnote{\texttt{http://www.atelierb.eu}}, 
ProB\footnote{\texttt{http://www.stups.uni-duesseldorf.de/ProB}} 
and RODIN\footnote{\texttt{http://www.rodintools.org}}. 

\eventB~\cite{Abrial2007} is an evolution of the B language to specify complex systems by using 
decomposition and event-based descriptions. In \eventB, specifications describe 
``closed'' event systems, in order to consider a system and its interactions with its environment as 
a whole. The behaviour is then modelled by events on the system. An event is defined by a guard, a 
blocking condition that ensures the consistency of the system if the event is executed, and an action 
described by GSL as in B. An event is of the form ANY $x,y,...$ WHERE $P(x,y,...,v,w,...)$ THEN
$S(x,y,...,v,w,...)$ END, where $x,y,...$ are local variables and $v,w,...$ are constants or state 
variables of the event system, 
predicate $P$ is the guard, and substitution $S$, the action. An event system may be refined. Refinement 
in \eventB not only refines data structures like in B, but also allows new events to be added. However,
only new concrete variables can be modified by new events. The state refinement is expressed, like in B, 
with a gluing invariant between the abstract state and the concrete state.

\subsection{ASTD}
\label{sec:astd}

\astd~\cite{FrGeLaFrSt:2008:revueISSE} is a formal graphical notation, which is an extension 
of Harel's Statecharts~\cite{statecharts} with process algebra operators. Each \astd type 
corresponds to either a hierarchical automaton or a process algebra operator like sequence, 
choice, Kleene closure, guard, synchronisation, choice and interleave quantification.
One of the main important features of \astd{s} is to allow parameterised instances 
and quantifications. Moreover, the graphical representation brings important means
for communicating with stakeholders and for validating the system model. This formal 
language has notably been used in the context of secure web services for security 
policy specification~\cite{milhauUMLFM2011,ar:mi-mde-10}.
For the sake of concision, we introduce only the \astd operators that
will be used in the case study: automaton, quantified parameterised synchronisation, Kleene closure and 
weak synchronisation. The complete operational semantics is in~\cite{MarcFrappier2008TR}. 

An \astd automaton is similar to a classical automaton, except that its states can be of 
any \astd type, and that its transition relation $\delta$ can refer to substates of automaton 
states. Hence, there are three kinds of arrows: local transition between two states $n_1$ 
and $n_2$ of the automaton, denoted by $(\lock, n_1, n_2)$; transition from $n_1$ to substate 
$n_{2_\flat}$ of $n_2$; and transition from 
substate $n_{1_\flat}$ of $n_1$ to $n_2$.
A transition can also be guarded or considered as final (\ie it is triggered only if its 
source state is final). Thus, a transition from $\delta$ is of the form $(t,\sigma,g,\finalTrans)$,
where $t$ denotes the arrow, $\sigma$ is the event, $g$ is the guard, and $\finalTrans$ is 
a boolean denoting whether the transition is final. For the sake of concision, history states 
are omitted here. A state of an automaton is of the form $ (\autStk,n,s) $ where $n$ is the name 
of its current state and $s$ is the current substate of the state. For example, rule $\autk_1$ describes the semantics of a local transition
\begin{center}
\AxiomC{$
  \begin{array}[b]{c}
  \delta((\lock, n_1, n_2),\sigma',g,\finalTrans)
    \mm
   \Psi
  \end{array}$}
\LeftLabel{$\autk_1$}
\UnaryInfC{$
   (\autStk, n_1, s)
      \xrightarrow{\sigma,\Gamma}
   (\autStk, n_2,\initk(\nu(n_2))) $}
\DisplayProof
\end{center}
Predicate $\Psi$ is a premiss which checks if the source state is final for final transitions, if the guard 
holds, and if the event received, noted $\sigma$, is equal, under the current transition environment $\Gamma$, 
to the event specified in the transition relation, noted $\sigma'$. Expression $\initk((\nu(n_2)))$ represents
the initial state of the \astd whose name is $n_2$. Thus, the target state of the transition is the initial 
state of the destination state in $\delta$.

Then we need Kleene closure: a Kleene closure \astd is an \astd that can be executed zero, one or more times. When the final state is reached, the \astd can restart.

The next \astd we consider is parameterised quantified synchronisation. The 
behaviour is defined as follows. Many \astd{s} are executed in parallel. For
each event whose label belongs to a synchronisation set $\Delta$, all
\astd{s} must execute this event at the same time; otherwise, they are 
executed by interleaving. The first requirement may be too strong to satisfy in some situations. In particular, if a quantified parameterised synchronisation is used to specify the behaviour of several entities in parallel, it would be very restrictive to prevent a large subset of entities from executing a synchronised event because some of them are not ready or can be considered as having stopped
their activity.

To take such cases into account, a weak synchronisation has been defined. A state of the \astd 
is then of the form $(\mbox{$\pitchfork$}\stdec,f)$, where $f$ maps an \astd state to each quantification
parameter value. In the type corresponding to this kind of \astd{s}, $\Delta$ represents the set 
of the actions that synchronise as above, and predicate $p$ characterises which instances of
the quantified \astd must synchronise. There are two inference rules: 
\begin{prooftree}
\AxiomC{$\alpha (\sigma)\notin\Delta$}
\AxiomC{$f(v)\xrightarrow[]{\text{$\sigma,([x:=v])\domres\Gamma$}}s'$}
\LeftLabel{$\pitchfork_1$}
\BinaryInfC{$(\mbox{$\pitchfork$}\stdec,f)\overset{\sigma,\Gamma}{\longrightarrow}(\mbox{$\pitchfork$}\stdec,f\ovl \{x\mapsto s'\})$}
\end{prooftree}
\begin{prooftree}
\AxiomC{$\alpha (\sigma)\in\Delta$}
\AxiomC{$\forall v \in T.((\lnot([x:=v]p)\land f(v) = f'(v))\mbox{$\lor$} (f(v)\xrightarrow[]{\text{$\sigma,([x:=v])$}} f'(v)))$}
\LeftLabel{$\pitchfork_2$}
\BinaryInfC{$(\mbox{$\pitchfork$}\stdec,f)\overset{\sigma,\Gamma}{\longrightarrow}(\mbox{$\pitchfork$}\stdec,f')$}
\end{prooftree}
Rule $\pitchfork_1$ is applied when there is no synchronisation. Rule $\pitchfork_2$ corresponds to the case with synchronisation: all the \astd{s} for which $p$ is true execute the event at the same time and the state 
of the other \astd{s} does not change.

\section{Overview of the approach}
\label{sec:overview}

\begin{figure}
  \centering
  \includegraphics[width=0.7\textwidth]{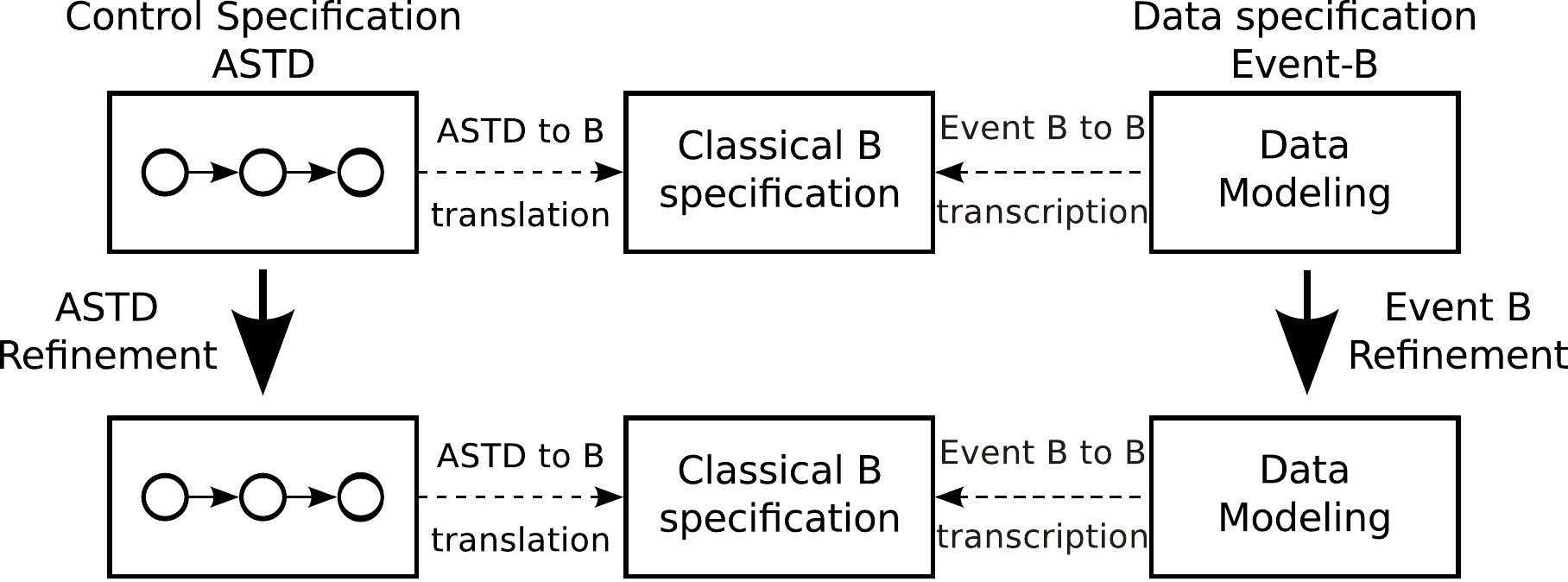}
  \caption{Methodology of the specification}
  \label{Meth}
\end{figure}

Our approach uses a coupling of the graphical \astd notation and \eventB to specify a system. The specification methodology is shown in Fig.~\ref{Meth}. A system can be viewed in two parts. The first part models the dynamic behaviour of the system, and is specified in \astd (box on the left in Fig.~\ref{Meth}). The second part focuses on data, and is described in \eventB (box on the right in Fig.~\ref{Meth}). Transitions constitute the link between the two parts: to each action label in \astd corresponds an event in \eventB. To ensure the global consistency of the system, the \astd and \eventB specifications are translated into classical B (middle box in Fig.~\ref{Meth}). As we will explain later, the classical B is only used for technical reasons.

\subsection{\astd specification.}

With graphical notations and process algebra operators, an \astd specification models the ordering of actions. Since formal notations are not always easy to understand, \astd provides a graphical visualisation which makes the model validation easier, while still remaining formal. Compared to Statecharts, the \astd language is based on process algebra operators, like quantified parameterised synchronisation, which allows to represent many processes in parallel.

\subsection{\eventB specification.}
\eventB specification contains an event for each action label declared in \astd. The \astd part just describes the ordering of actions. In the \eventB part we specify the effects  on the data model of each \astd action. Static properties like safety and typing constraints are specified by means of \eventB invariants. Sometimes we need temporal properties which are not supported by the \eventB notation. In that case, we encode these temporal properties by using theorems. Rodin tools are used to generate and prove the proof obligations associated to invariant preservation and additional theorems.

\subsection{B specification.}
The classical B specification contains two B machines. The first one is the translation of the \astd specification, the second one is a transcription of the \eventB specification.

The \astd to B translation can be summed up as follows. \astd states are encoded by B state variables. To every \astd action label corresponds a B operation. Its precondition checks that the state variable is in the initial state of the \astd transition. Its postcondition assigns to the state variable the final state of the transition. Moreover, to link the resulting B operation with the data model, we would like to execute the events defined in the \eventB part during the transition. But technically, a B operation cannot call an \eventB event. That is why we also have to translate the events into B operations.

For the translation of the \eventB machine, variables and typing invariants remain unchanged. Events are rewritten into B operations: their guards are simply changed into preconditions and their postconditions remain identical. Grouping the two parts together in one unique B specification allows the global consistency (one horizontal level in Fig.~\ref{Meth}) of the system to be proved: when we call an operation in B, the generated proof obligation checks that the precondition of the called operation is true before executing it. To prove the calling proof obligations, invariants are added in the B machine that is the translation of the \astd. These invariants link the variables of the \eventB description and the variables that encode the states of the \astd.

 \eventB provides the expressiveness and the refinement relation required for the system to be modelled, but it lacks some modularity features. There exist theoretical foundations for modularity in \eventB~\cite{DBLP:conf/asm/IliasovTLRVIL10}, but in practice, they are not yet supported by existing tools. B is then used for technical reasons.  

\subsection{Refinement of the model.}

The methodology uses two refinements. On one side we refine the \astd specification (left refinement arrow in Fig.~\ref{Meth}), on the other side we refine the data specification in \eventB (right refinement arrow in Fig.~\ref{Meth}).

A first definition of \astd refinement is proposed in~\cite{FACS-journal-2014}. This refinement definition requires the traces to be preserved and three generic application patterns are described. By trying to apply this \astd refinement relation on the case study described here, we realise that it is too restrictive. Consequently we have introduced new patterns that weaken the original definition but preserve behaviour consistency: the properties that are true in a state of the abstract \astd specification have to be preserved in the corresponding state of the concrete specification. This new refinement definition is detailed in section~\ref{sec:secondRef}.

The \eventB part of the specification is refined using \eventB definition of refinement. The proof obligations are automatically generated by the RODIN tool. The \eventB refinement guarantees the preservation of the invariants in the data specification. This refinement definition is one of the reasons why we chose \eventB to specify the data part of the system: The classical B refinement does not allow events to be added, while \astd refinement allows new transitions.

\section{Case Study}
\label{sec:casestudy}

Our \eventB /\astd coupling is used to specify a train system, more precisely a CBTC-like train controller. CBTC is an automatic train control system based on communications between two subsystems. The track controller manages the entire track where the trains are moving. An on-board controller drives each train. At the most abstract level, the specification consists of trains moving independently. The aim is to define a two-part-system: a part of the system drives the trains, another part manages all the trains on the track. Informally, the property we want to prove for the system is the absence of collision.

Using formal methods to specify railway systems has already been done in the litterature. Ferrari and al.~\cite{Ferrari2014} use semi-formal methods to specify a complete CBTC specification. They define a methodology in a software product line approach. Many articles describe the use of formal methods to specify interlocking system: Abrial specifies it with \eventB in the \eventB book~\cite{Abrial2007}, James \emph{et al.} with CSP$\|$B~\cite{James}. Silva uses \eventB too to specify a train system in his PhD thesis~\cite{Silva2012}. Our specification starts from a more abstract global view of the system and we refine it into the specification of two controllers.  The other approaches directly specify the behaviour of each controller, without considering their global composition as we do.

\subsection{Modelling Choices}

At the most abstract level, we consider a unique track on which a set of trains are moving in the same direction. This is realistic since, if there are trains in the opposite direction, they are blocked at switches. Furthermore, all issues concerning switches, interlocking, etc... are not considered in this paper. They can be dealt with in subsequent refinement steps. Consequently, we check only the absence of front-to-rear collision

We define a set named $\mathit{TRACK}$ on which there is a total, strict order relation (irreflexive, transitive and asymetric) called $\mathit{is\_behind}$. $x_1 \mapsto x_2 \in \mathit{is\_behind}$ means that the element $x_1$ is behind the element $x_2$ on the track.

In the following, we present four levels of specification. In the first one, the trains are moving independently, the second one introduces a controller for each train, the third one groups the controllers together into one single control operation and the last one splits the variables between two entities, preparing for the decomposition.

\subsection{First Specification}

\label{sec:firstSpec}



\begin{figure}
\centering
\includegraphics[width=0.5\textwidth]{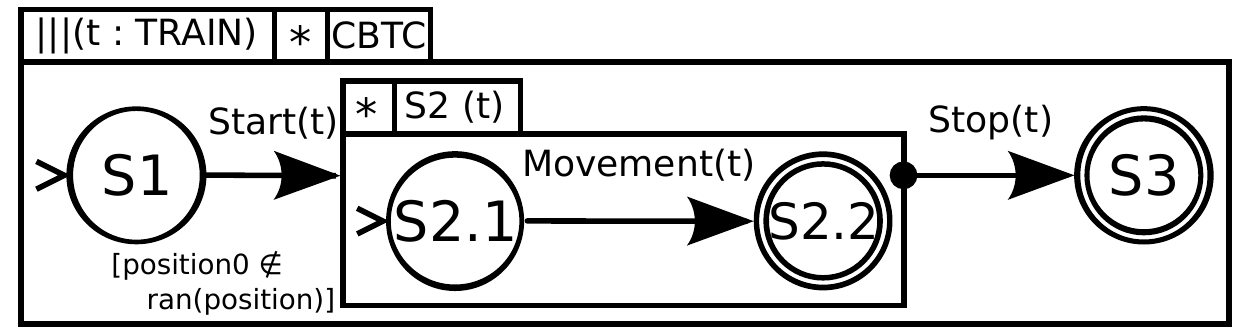}
\caption{First \astd Specification}
\label{CBTC_N2}
\end{figure}

We start by specifying the system behaviour in \astd (in Fig.~\ref{CBTC_N2}). The fact that we have many trains is represented by the quantified interleaving operator. Each train can start, move and stop. $S2(t)$ is an \astd of type Kleene closure: this means that the body of $S2(t)$ (\ie the $movement$ transition) can be performed zero, one or more times.  When it is stopped, a train can restart thanks to the Kleene closure.

Since we assume that the trains are moving in the same direction on one track, the non-collision property can be expressed by two predicates:

\noindent\begin{align} \label{Prop:prop1}
\forall (t_1,t_2).(&t_1\in \mathit{TRAIN}\ \wedge t_2 \in \mathit{TRAIN} \wedge t_1 \neq t_2 \Rightarrow \nonumber \\
 &\hspace{1cm} \mathit{position}(t_1) \neq \mathit{position}(t_2))
\end{align}
Predicate~(\ref{Prop:prop1}) means that the position of two distinct trains are different. Another predicate is needed to express that a train cannot jump over another train.
\begin{align} \label{Prop:prop2}
\forall (t_1,t_2) .(&t_1 \in \mathit{TRAIN}\ \wedge t_2 \in \mathit{TRAIN} \wedge \mathit{position}(t_1) \mapsto \mathit{position}(t_2) \in \mathit{is\_behind} \Rightarrow \nonumber \\
& \hspace{1cm}\mathsf{X}(\mathit{position}(t_1) \mapsto \mathit{position}(t_2) \in \mathit{is\_behind}))
\end{align}
Predicate~(\ref{Prop:prop2}) checks that the order of trains does not change. Symbol $\mathsf{X}$ denotes the "next" operator from temporal logic~\cite{pnueli81}.

To express these predicates in \eventB we introduce a set $\mathit{TRAIN}$ and a state variable $\mathit{position}$ which is a partial function from $\mathit{TRAIN}$ to $\mathit{TRACK}$ ($\mathit{position} \in \mathit{TRAIN} \pfun \mathit{TRACK}$). Variable $\mathit{position}$ is set when the train starts and until it stops. The EventB event, corresponding to the movement action, that acts on the data is called $\mathit{movement\_act}$. It updates the variable such as (see Figure~\ref{evb:movement}):
\begin{itemize}
\item the new position is different from the positions of the other trains;
\item the new position stays behind the position of the trains that were located before;
\item the new position cannot be located behind the old position.
\end{itemize}

\begin{figure}
  \begin{description}
    \EVT {\textit{movement\_act}}
    \begin{description}
      \AnyPrm
      \begin{description}
      \Item{tt }
      \end{description}
      \WhereGrd
      \begin{description}
        \nItem{ gu1 }{ tt \in  TRAIN }
        \nItem{ gu2 }{ tt \in  \dom(position) }
      \end{description}
      \ThenAct
      \begin{description}
        \nItem{ act1 }{ \textit{position} :|   (\exists \textit{pp} \qdot (\textit{pp} \in  \textit{TRACK} \land\textit{position'} = \textit{position} \ovl  \{ \textit{tt}\mapsto  \textit{pp}\}  \land \\
          \hspace*{1 cm}  (\forall t_2\qdot (t_2 \in  \dom(\textit{position'}) \land  t_2 \neq  \textit{tt}   \limp  \textit{pp} \neq  \textit{position'}(t_2))) \land\\
          \hspace*{1 cm}  (\forall t_2\qdot (t_2 \in  \dom(\textit{position}) \land \textit{position}(tt) \mapsto \textit{position}(t_2) \in  \textit{is\_behind} \limp\\
          \hspace*{2 cm} \textit{pp} \mapsto  \textit{position}(t_2) \in \textit{is\_behind})) \land\\
          \hspace*{1 cm}  (\textit{pp} = \textit{position}(\textit{tt}) \lor \textit{position}(\textit{tt}) \mapsto  \textit{pp} \in \textit{is\_behind}))) }
      \end{description}
      \EndAct
    \end{description}
  \end{description}
  \caption{First Specification: The \textit{movement\_act} event}
  \label{evb:movement}
\end{figure}

Predicate~(\ref{Prop:prop1}) is directly defined as an invariant of the data specification. Predicate~(\ref{Prop:prop2}) uses an operator coming from temporal logic and cannot be model checked by ProB. To avoid this issue, we translate temporal logic predicate~(\ref{Prop:prop2}) into assertions on the states, written as \eventB theorems. An \eventB theorem is an assertion that has to be proved with the invariants of the machine. A theorem is written for each event of the machine. For example, the theorem corresponding to the $\mathit{movement\_act}$ event checks that for all trains $\mathit{train_1}$, $\mathit{train_2}$ and $\mathit{train_3}$ such that $\mathit{train_1} \neq \mathit{train_2}$, (a) if $\mathit{train_1}$ is behind $\mathit{train_2}$, (b) if the precondition of the $\mathit{movement\_act}$ operation is true for $\mathit{train_3}$, (c) if we execute $\mathit{movement\_act}(\mathit{train_3})$ then $\mathit{train_1}$ stays behind $train_2$. Note that this theorem includes the three possible cases: $\mathit{train_3} = \mathit{train_1}$, $\mathit{train_3} = \mathit{train_2}$ and ($\mathit{train_3} \neq \mathit{train_1}$ and $\mathit{train_3} \neq \mathit{train_2}$)

\astd and \eventB specification are then translated into classical B. We detail the example of the $\mathit{movement}$ action label. In the \astd translation, a $\mathit{movement}$ operation is created. The precondition of the operation verifies that the state variable is the initial state of the precondition (state 2.1 of \astd $S2(t)$ in Fig.~\ref{CBTC_N2}). The postcondition assigns the final state of the transition to the state variable (state 2.2 of \astd $S2(t)$ in Fig.~\ref{CBTC_N2}) and calls the translation of the \eventB $\mathit{movement\_act}$ operation. The operation is

\vspace{-0,8cm}

\begin{align*}
  &\mathit{movement}(t) =\\
  &\mathbf{PRE}\ \mathit{State\_S2}(t) = S2.1\\
  &\mathbf{THEN}\ \mathit{State\_S2}(t) \bcmeq S2.2 || \mathit{movement\_act}(t)\\
  &\mathbf{END}.\\
\end{align*}

\vspace{-0,8cm}

\noindent The translation of $\mathit{movement\_act}$ operation is

\vspace{-0,8cm}

\begin{align*}
  &\mathit{movement\_act}(t) =\\
  &\mathbf{PRE}\ t \in \dom(\mathit{position})\\
  &\mathbf{THEN}\ \mathit{position} :|(P )\\
  &\textbf{END}\\
\end{align*}

\vspace{-0,8cm}

\noindent where $P$ is the same predicate as in the \color{blue} act1 \color{black} part of the \eventB postcondition of $\mathit{movement\_act}$.

In the data specification we have already proved that the invariant is preserved if the guards of the events are true. In the unifying machine, generated proof obligations require to prove the precondition (hence the guard of the event) of the called operation $\mathit{movement\_act}$ ($t \in \dom(\mathit{position})$) to be true when the operation is called (\ie $\mathit{State\_S2}(t) = s1.1$). These two proofs imply that the invariant is always preserved. This justifies the need of a unifying machine for proving the horizontal consistency.

\subsection{First Refinement}

\label{sec:firstRef}

At the abstract level, the $\mathit{movement\_act}$ operation describes the properties that the new position must satisfy. In this refinement, we show how such a position can be chosen. A computing operation is added for each train. It computes a limit for the train, called $\mathit{mal}$ (\textit{Movement Authority Limit}), given all the positions of the other trains. The $\mathit{movement\_act}$ operation now updates the position such that the train cannot overtake the limit.

\begin{figure*}
\centerline{\includegraphics[width=0.8\textwidth]{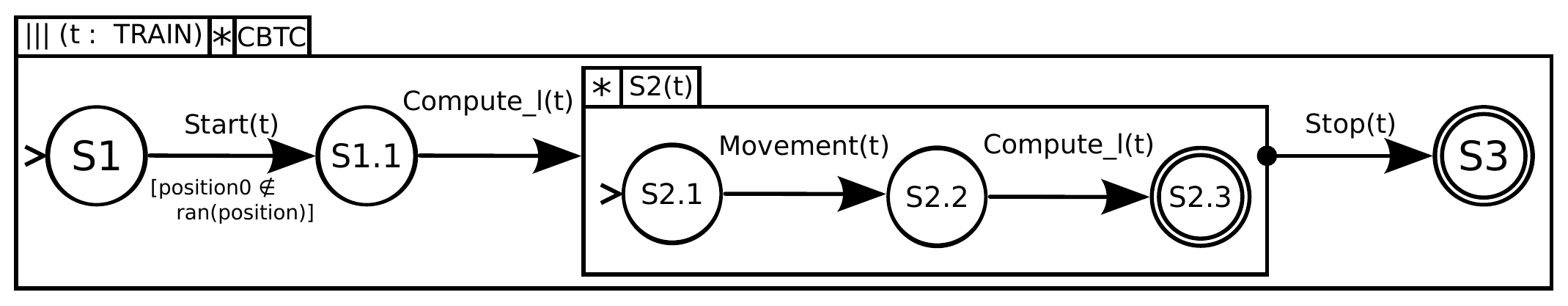}}
\caption{First Refinement - ASTD Specification}
\label{CBTC_N3}
\end{figure*}

In the \astd specification, the action label $\mathit{compute\_l}$ that computes the limit is added twice. Just after a train starts, the system computes a $\mathit{mal}$. Each time a train has moved, the system has to compute a new $\mathit{mal}$. The new ASTD specification is in Fig.~\ref{CBTC_N3}. The refinement is proved using the refinement definition given in~\cite{FACS-journal-2014}.

Since a new action label is added in the \astd, we need to describe its effects on the data model in \eventB. It computes a limit for the train. This limit is the $\mathit{mal}$ variable of the \eventB specification. The invariant properties associated to $\mathit{mal}$ are:

\begin{align} \label{Prop:prop3}
\forall (t_1,t_2).(&t_1 \in \mathit{TRAIN} \wedge t_2 \in \mathit{TRAIN} \wedge \nonumber\mathit{position}(t_1) \mapsto \mathit{position}(t_2) \in \mathit{is\_behind} \Rightarrow\nonumber\\
&\hspace{1cm}\mathit{mal}(t_1) \mapsto \mathit{position}(t_2) \in \mathit{is\_behind}
\end{align}

\vspace{-0,8cm}

\begin{align} \label{Prop:prop4}
\forall t.(&t \in \mathit{TRAIN}\Rightarrow (\mathit{position}(t) \mapsto \mathit{mal}(t) \in \mathit{is\_behind}\ \vee \mathit{position}(t) = \mathit{mal}(t)))
\end{align}
Predicate~(\ref{Prop:prop3}) checks that the $\mathit{mal}$ of a train is always located behind the trains that are in front of it. Predicate~(\ref{Prop:prop4}) checks that a train cannot overtake its limit.

The $\mathit{compute\_l\_act}$ operation (see Figure~\ref{evb:computeL}) computes the limit for a train $t$ such that: (a) the new $\mathit{mal}(t)$ is in front of or equal to current $\mathit{position}(t)$, (b) for all trains $t_2$ whose $\mathit{position}(t_2)$ is in front of $\mathit{position}(t)$, the new $\mathit{mal}(t)$ is behind $\mathit{position}(t_2)$.

\begin{figure}
  \begin{description}
    \EVT {\textit{compute\_l\_act}}
    \begin{description}
      \AnyPrm
      \begin{description}
      \Item{tt }
      \end{description}
      \WhereGrd
      \begin{description}
	\nItem{ gu1 }{ \textit{tt} \in  \textit{TRAIN} }
	\nItem{ gu2 }{ \textit{tt} \in  dom(\textit{position}) }
      \end{description}
      \ThenAct
      \begin{description}
	\nItem{ act1 }{ \textit{mal} :|  (\exists \textit{mm}\qdot (\textit{mm} \in  \textit{TRACK} \land\\
          \hspace*{1cm}  \forall t_2\qdot (t_2 \in  \textit{TRAIN} \land t_2 \in  \dom(\textit{position}) \land\\
          \hspace*{1cm}  \textit{position}(\textit{tt}) \mapsto  \textit{position}(t_2) \in  \textit{is\_behind} \limp \textit{mm} \mapsto  \textit{position}(t_2) \in  \textit{is\_behind}) \land\\
          \hspace*{1cm} (\textit{position}(\textit{tt}) \mapsto  \textit{mm} \in  \textit{is\_behind} \lor \textit{position}(\textit{tt}) = \textit{mm}) \land\\
          \hspace*{1 cm}  \textit{mal'} = \textit{mal} \ovl  \{ \textit{tt} \mapsto  \textit{mm}\} )) }
      \end{description}
      \EndAct
    \end{description}
  \end{description}
  \caption{First Refinement: The \textit{compute\_l\_act} Event}
  \label{evb:computeL}
\end{figure}

Since we compute limit for the trains, we modify the $movement\_act$ operation (Figure~\ref{evb:movement2}) such that the new position of a train $t$ is chosen depending on $mal(t)$: the new position $pp$ is between the old $position(t)$ and $mal(t)$.

\begin{figure}
  \begin{description}
    \EVT {\textit{movement\_act}}
    \REF {\textit{movement\_act}}
    \begin{description}
      \AnyPrm
      \begin{description}
      \Item{tt }
      \end{description}
      \WhereGrd
      \begin{description}
	\nItem{ gu1 }{ \textit{tt} \in  \textit{TRAIN} }
	\nItem{ gu2 }{ \textit{tt} \in  \dom(\textit{position}) }
	\nItem{ gu3 }{ \textit{tt} \in  \dom(\textit{mal}) }
      \end{description}
      \ThenAct
      \begin{description}
	\nItem{ act1 }{ \textit{position} :|  (\exists \textit{pp}\qdot ( \textit{pp} \in  \textit{TRACK} \land \textit{position}' = \textit{position} \ovl  \{  \textit{tt} \mapsto  \textit{pp} \}  \land\\
          \hspace*{1cm}  (\textit{position}(\textit{tt}) = \textit{pp} \lor (\textit{position}(\textit{tt}) \mapsto  \textit{pp}) \in  \textit{is\_behind}) \land\\
          \hspace*{1cm}  (\textit{pp} = \textit{mal}(\textit{tt}) \lor  (\textit{pp} \mapsto  \textit{mal}(\textit{tt})) \in  \textit{is\_behind}))) }
      \end{description}
      \EndAct
    \end{description}
  \end{description}
  \caption{First Refinement: The \textit{movement\_act} Event}
  \label{evb:movement2}
\end{figure}

Predicates~(\ref{Prop:prop1}) and~(\ref{Prop:prop2}) are preserved by refinement. Predicate~(\ref{Prop:prop1}) is obvious because of invariant preservation by refinement. Predicate~(\ref{Prop:prop2}) was translated in terms of postconditions and refinement guarantees the compatibility of postconditions of refined events.

\subsection{Second Refinement}
\label{sec:secondRef}

\begin{figure*}
\centerline{\includegraphics[width=0.8\textwidth]{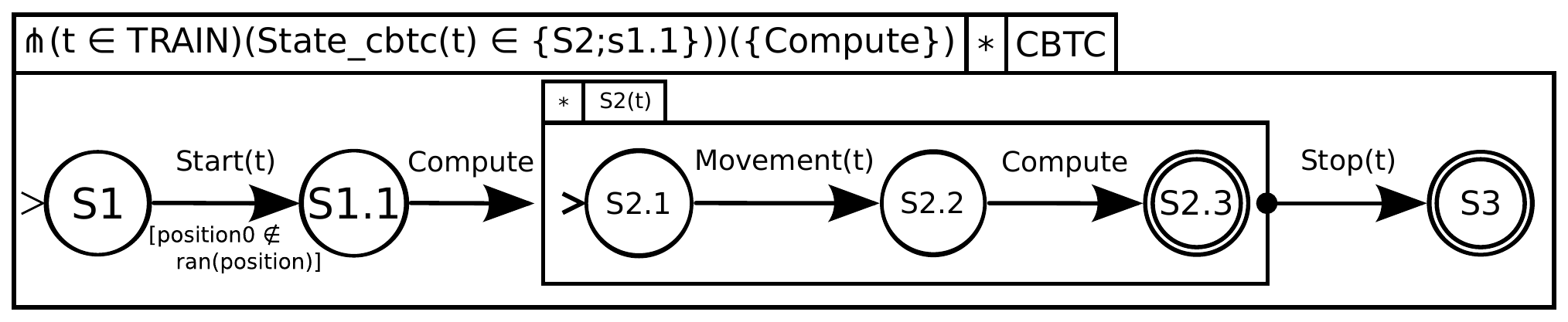}}
\caption{Second Refinement - ASTD Specification}
\label{CBTC_N4}
\end{figure*}

In this level of refinement, all the local computing operations $\mathit{compute\_l}(t)$ are grouped into one global $\mathit{compute}$ operation. It is synchronised for all the started trains ($\pitchfork$ operator). The new specification is depicted in Fig.~\ref{CBTC_N4}.

This refinement transforms an interleaving \astd $|||(t\in \mathit{TRAIN}) A(t)$ into a synchronised \astd $\pitchfork (t\in \mathit{TRAIN}) C(t)$. The set of accepted traces is restricted. But we want to preserve behaviour consistency: For each train $t_n \in \mathit{TRAIN}$ the \astd $C(t_n)$ is a refinement of the \astd $A(t_n)$ according to the definition proposed in~\cite{FACS-journal-2014}.  It means that if the global set of traces accepted by the \astd specification is reduced, the local behaviour of each entities is preserved.

Since we change the local $\mathit{compute\_l}$ operation into a global one, we need to define this operation in the data specification. This operation computes for all trains a limit such that the limit respects the invariants defined in section~\ref{sec:firstRef}. The specification of $\mathit{compute\_act}$ is shown on Figure~\ref{evb:compute}.

\begin{figure}
  \begin{description}

    \EVT {\textit{compute\_act}}
    \begin{description}
      \BeginAct
      \begin{description}
	\nItem{ act1 }{ mal :|  ((dom(\textit{mal'}) = \dom(\textit{position}) \land \textit{mal}' \in  \textit{TRAIN} \pfun  \textit{TRACK} \land\\
          \hspace*{1cm}  (\forall t_1,t_2\qdot (t_1 \in  \dom(\textit{position}) \land t_2 \in  \dom(\textit{position}) \land \\
          \hspace*{1,2cm} \textit{position}(t_1) \mapsto    \textit{position}(t_2) \in  \textit{is\_behind} \limp\\
          \hspace*{1,2cm} \textit{mal}'(t_1) \mapsto  \textit{position}(t_2) \in  \textit{is\_behind})) \land\\
          \hspace*{0,8cm}  (\forall \textit{tt}\qdot (\textit{tt}  \in  dom(\textit{mal}') \limp  (\textit{position}(\textit{tt}) \mapsto  \textit{mal}'(\textit{tt}) \in  \textit{is\_behind}\lor\\
          \hspace*{1cm}\textit{mal}'(\textit{tt}) = \textit{position}(\textit{tt})))))) }
      \end{description}
      \EndAct
    \end{description}

  \end{description}
  \caption{Second Refinement: The \textit{compute\_act} Event}
  \label{evb:compute}
\end{figure}

Proving this refinement is not possible in \eventB: a set of local events cannot be refined by a global one. To prove the consistency of our specification, we proved that executing $\mathit{compute\_act}$ operation is equivalent to execute any sequence of $\mathit{compute\_l\_act}$ (which means an interleaving of $\mathit{compute\_l\_act}$).

To prove the refinement, events are expressed as relations between the state of a variable before and after executing the event. We write $\mathit{Rel}_{\mathit{Ev}}$ the relation for an event $\mathit{Ev}$ and $\mathit{Rel}_{\mathit{Ev}}(t)$ if the event has a parameter. We proved that (a) for all $t_1$ and $t_2$, $t_1$ and $t_2$ being  in $\mathit{TRAIN}$, $\mathit{Rel}_{\mathit{Compute\_l\_act}}(t_1);\mathit{Rel}_{\mathit{Compute\_l\_act}}(t_2) = \mathit{Rel}_{\mathit{Compute\_l\_act}}(t_2);\mathit{Rel}_{\mathit{Compute\_l\_act}(t_1)}$ which means that for all couples of trains, the order in which we execute $\mathit{Compute\_l\_act}$ event does not change the result. Using (a) and by induction on the set of trains, we proved that all the sequences of $\mathit{Compute\_l\_act}$ execution are equivalent. Finally, we proved that executing an arbitrary sequence of $\mathit{Compute\_l\_act}$ is equivalent to execute $\mathit{Compute\_act}$. This implies that executing $\mathit{Compute\_act}$ is equivalent to execute an interleaving of $\mathit{Compute\_l\_act}$ event.

\subsection{Third Refinement}
\label{sec:thirdRef}

We want our system to have two subsystems. The on-board system drives the trains. It modifies variable $\mathit{position}$, using variable $\mathit{mal}$. The track system manages the entire subsystem. It computes variable $\mathit{mal}$ using variable $\mathit{position}$. To share the variables between two subsystems, communications are introduced.

Each variable of the second refinement is refined by two variables: one variable called track variable is used and modified by the track controller and the other one called on-board variable is used and modified by the on-board controller. The gluing invariant is $\mathit{mal} = \mathit{track\_mal} \land \mathit{position} = \mathit{on\_board\_position}$.

In the data specification, variable $\mathit{mal}$ is replaced by $\mathit{track\_mal}$ in the operations that modifies it, and variable $\mathit{position}$ is replaced by $\mathit{on\_board\_position}$. The rest of the specification remains almost unchanged: some guards are added to prove the \eventB refinement. A communication operation $\mathit{CommBT\_act}$ (resp. $\mathit{CommTB\_act}$) send variable $\mathit{on\_board\_position}$ (resp. $\mathit{track\_mal}$) from the board to the track (resp. from the track to the board). The specification of the $\mathit{CommBT\_act}$ is shown in Figure~\ref{evb:commBT}

\begin{figure}
  \begin{description}
    \EVT {commBT\_act}
    \begin{description}
      \AnyPrm
      \begin{description}
      \Item{tt }
      \end{description}
      \WhereGrd
      \begin{description}
        \nItem{ gu1 }{ tt \in  TRAIN }
        \nItem{ gu2 }{ tt \in  dom(on\_board\_position) }
      \end{description}
      \ThenAct
      \begin{description}
        \nItem{ act1 }{ track\_position(tt) :=  on\_board\_position(tt) }
      \end{description}
      \EndAct
    \end{description}
  \end{description}
  \caption{Third Refinement: The \textit{commBT\_act} event}
  \label{evb:commBT}
\end{figure}

\begin{figure}
  \begin{center}
    \includegraphics[width=\textwidth]{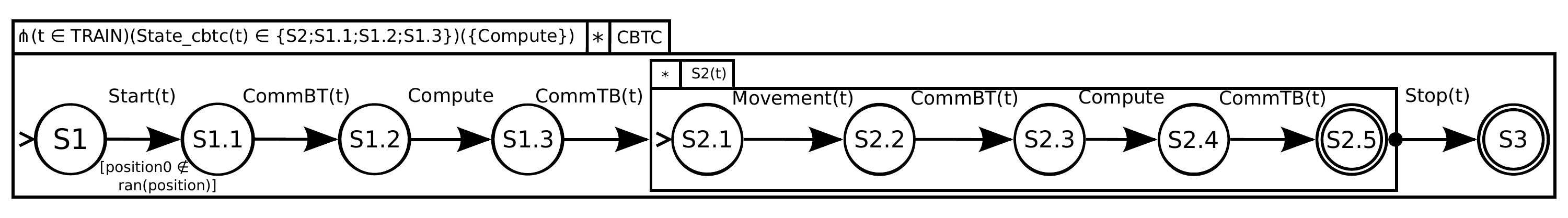}
  \end{center}
  \caption{Train System \astd Specification: Fourth Refinement}
  \label{astd:N5}
\end{figure}

In the control specification, after each operation that modifies variable $\mathit{on\_board\_position}$ (the operation $\mathit{Start}$ and $\mathit{Movement}$), a communication operation from the board to the track ($\mathit{CommBT}$) is added. Dually, a communication operation from the track to the board ($\mathit{CommTB}$) is added after the $\mathit{Compute}$ operation. The \astd specification can be seen in Figure~\ref{astd:N5}.

The \astd refinement follows the refinement definition of~\cite{FACS-journal-2014}. The refinement of the data specification follows the \eventB definition of the refinement.

\subsection{Discussion about the specification}
\label{sec:discussion}

In this section, we want to evaluate the efficency of our specification method. For this purpose, a benchmark specification was specified. First, we sum up our case study. Then the benchmark specification is explained. Finally, the specifications are compared.

\paragraph{Summary of the case study}

In this paper, four levels of specification are presented out of six in the entire case study. The rest of the case study is described in a technical report~\cite{Fayolle2014}. The table in Figure~\ref{tab:sumup} sums up the statistics of this case study. It contains the number of lines of specification and the number of proof obligations. The specifications presented in this paper are level 2 to 5. Our aim is to evaluate the efficiency of the \astd langage. Thus we only present the statistics about the control specification. The second column is the number of lines in the generated B specification. The last four columns are the number of generated proof obligations. The first column is the total number of proof obligations, the second and third ones are respectivelly the number of automatically and manually proved obligations. The Atelier B tool allows one to save the user proof tactics. Since some proof obligations are similar, some tactics can prove many proof obligations. The last column contains the number of proof tactics that are sufficient to prove all proof obligations.

\begin{figure}
  \centering
  \begin{tabular}{|c|c|c|c|c|c|c|}
    \hline
    \multirow{2}{*}{{\begin{tabular}[c]{@{}c@{}}\\Specification\\ Level\end{tabular}} } &
    \multirow{2}{*}{{\begin{tabular}[c]{@{}c@{}}Number of lines:\\Generated B\\Specification\end{tabular}}} &
    \multicolumn{4}{c|}{Number of Proof Obligations}\\ \cline{3-6} 
    & & Total & \begin{tabular}[c]{@{}c@{}}Automatically\\ Proved\end{tabular} & \begin{tabular}[c]{@{}c@{}}Manually\\Proved\end{tabular} & \begin{tabular}[c]{@{}c@{}}Number of\\ User Tactics\end{tabular} \\ \hline
                Level 1 & 98   & 30   & 25   & 5   & 5  \\ \hline 
                Level 2 & 145  & 265  & 225  & 40  & 4  \\ \hline
                Level 3 & 185  & 342  & 252  & 90  & 14 \\ \hline
                Level 4 & 201  & 391  & 239  & 152 & 15 \\ \hline
                Level 5 & 277  & 871  & 337  & 534 & 58 \\ \hline 
                \begin{tabular}[c]{@{}c@{}}Benchmark\\ Specification\\\end{tabular} & 121 & 149 & 16 & 133 & 13\\ \hline
                Level 6 & 422  & 2710 & 1989 & 721 & 60 \\ \hline
  \end{tabular}
  \caption{Size of the case study}
  \label{tab:sumup}
\end{figure}

\paragraph{The benchmark specification}

To see the advantages and drawbacks of our methodology, a benchmark B specification was written following the traditional development process. This benchmark specification is a specification of the control specification that we have manually derived from the \astd specification. The \astd specification is used as a user requirement.

One state variable is added for each transition label of the \astd specification. This state variable contains the instance of trains for which the transition is enabled. The precondition of an operation verifies if the transition is enabled. The post condition removes the instance for the disabled operations, and adds it for the enable operation. For example, the $\mathit{movement}$ operation is shown in Figure~\ref{evb:movement3}. It means that the $\mathit{Movement}$ transition can be executed for the train $tt$ if it is in the set $\mathit{movement\_enabled}$. After this operation, neither $\mathit{Movement}$ nor $\mathit{Stop}$ operations are enabled, $\mathit{comm\_BT}$ becomes enabled. The benchmark specification contains 121 lines and generates 149 proof obligations, 133 are proved manually with 13 user tactics.

\begin{figure}
  \begin{align*}
    	&\mathit{Movement}(\mathit{tt}) =\\
	&\PRE \\
        &\hspace{1cm}   tt \in \mathit{TRAIN} \land\\
	&\hspace{1cm}	tt \in \mathit{movement\_enabled}\\
	&\THEN \\
        &\hspace{1cm}   \mathit{comm\_BT\_enabled} := \mathit{comm\_BT\_enabled} \cup \{\mathit{tt}\} ||\\
	&\hspace{1cm}	\mathit{movement\_enabled} := \mathit{movement\_enabled} - \{\mathit{tt}\} ||\\
	&\hspace{1cm}	\mathit{stop\_enabled} := \mathit{stop\_enabled}-\{\mathit{tt}\} ||\\
	&\hspace{1cm}	\mathit{Movement\_act}(\mathit{tt})\\
	&\ENDB;
  \end{align*}
  \caption{The benchmark specification of the movement transition}
  \label{evb:movement3}
\end{figure}

\paragraph{Comparison and discussion}

We compare the level 5 of our case study with the benchmark specification. Level 5 required to manually prove 534 obligations which were discharged with 58 tactics. The benchmark specification is shorter (121 lines vs 277) and generates less proof obligations (149 vs 871, 13 tactics required vs 58) than our specification. This comes from the fact that the control specification is automatically generated. As a comparison, the $\textit{Movement}$ operation in the generated B code contains 4 levels of \SELECT substitution, which generate a lot of different proof cases. The specification of this operation is not given here for the sake of concision; it is about 30 lines long.

On one hand, the benchmark specification is shorter and easier to prove, but it is almost impossible to be sure it respects the behaviour of the \astd specification without proving it. On the other hand, the \astd/B specification is automatically translated from the \astd specification: thus we are sure it respects the \astd specification but the resulting specification is a little hard to prove. We also try to prove the equivalence between the \astd/B specification and the benchmark specification. This proof is about as hard as proving the entire \astd/B specification, and mistakes were found which shows the necessity of the proof of equivalence.

This comparison provides some evidence that the \astd langage is a good, formal and easy to read langage to express user requirements. It can be used in a formal specification methodology, but the resulting specification is hard to prove and a little complex. In the future, we plan to improve the specification methodology to reduce the complexity of the generating specification and to help the user with the proof work.

\section{Conclusion}
\label{sec:conclusion}

\subsection{Related work.}
\label{sec:relatedw}

There exist several methods that combine state-based specifications with process algebras.
In csp2B~\cite{csp2b99}, a CSP specification is translated into a B machine. Only a subset 
of CSP operators are allowed for semantics reasons and there are some restrictions
on the use of synchronisation and interleaving operators. Contrary to our approach, the basic idea 
of csp2B is to use one of the two languages (\eg B) as a central support for verification and 
refinement. 

CSP $||$ B~\cite{treharne99using,DBLP:journals/fac/SchneiderT05}
consists in associating B machines with CSP controllers, such that B operations are constrained 
by the CSP process. Communications between B machines are modelled through their respective CSP 
controllers. The approach provides the sufficient conditions to verify the consistency of the 
whole specification. This work is probably the closest to our approach. Intrinsically, the 
decomposition of the system into several pairs B machine/CSP controller implies a clear 
separation between data model and behavioural description. However, it also induces a very 
constrained organisation of the model, which may be not straightforward for some systems.
Moreover, compared to \astd, expressiveness of CSP controllers is limited (see~\cite{FrGeLaFrSt:2008:revueISSE} 
for a comparison between \astd and CSP).  

Circus~\cite{circussem,oliveira2005refinement} is a formal language based on Z~\cite{usingz} 
and CSP, which integrates the refinement calculus of~\cite{actsysback}. The semantics is 
inspired by the unifying theory of programming~\cite{hoarehe}. The key idea is to distinguish
state transitions from the communications of the main action system that represents the 
behaviour of the system. CSP roughly offers the same process algebra as \astd but without a graphical view.
Moreover our data refinement is based on \eventB.

Finally iUML-B combines \eventB with UML. A state machine or a class diagram can be
added to an \eventB machine. Those diagrams are translated into \eventB specifications.
It does not support the quantified synchronisation that
is needed in our case study. Moreover, there is no formal semantics for now.

Let us now focus on refinement. Introducing extra operations is one of the main issues. 
\eventB refinement~\cite{Abrial2007} allows that, as opposed to classical B. In several 
event-based formal languages like CSP, a denotational semantics is defined~\cite{roscoe97}: 
refinement is then based on inclusions of sets representing observations of the system behaviour. 
The simplest observation is to consider all the sequences of operations that the system can perform; 
this corresponds to the \emph{traces} model. Contrary to \astd refinement, such an inclusion of sets 
of traces allows loss of traces during refinement. Moreover, this semantics does not allow new operations 
to be added. There are also some minor differences in the other observation models (failures, 
divergences) such that \astd allows definition of extra operations during refinement 
(see~\cite{FACS-journal-2014} for a more detailed comparison between CSP and \astd refinements).
Several work~\cite{DBLP:conf/zum/DerrickW05,DBLP:journals/scp/SchneiderT11,DBLP:journals/fac/Boiten14} 
deal with the consequences of the introduction of extra operations on the CSP, B or Z refinement 
semantics. Compatibility in such couplings requires observations on the input/output operations 
in the models used for refinement. This is consistent with respect to the \astd refinement definition.

\subsection{Lessons learnt and perspectives.}

Our main objective is to formalise the process of transforming informal, high-level user requirements  into
a global abstract formal specification by stepwise refinement, as advocated in the Event-B method.
Through a railway CBTC-like case study, we have shown how to model with two complementary formal languages, 
\eg \astd and \eventB. Then, we have explored the complementarity and consistency between \astd and B-like refinements.

By adopting a specification in two parts, there is a clear separation between data aspects and behavioural 
description. The refinement steps illustrated in the case study show the interest of considering both 
refinements. For instance, the first refinement step is \astd-oriented and brings more details 
in the event ordering for describing a specific process. When the state space must be updated, then 
the \eventB part will be refined. Such refinements are focused on one part of the model, and the idea 
is to propagate on the other part when some extra specifications are required (\eg when a new event 
is inserted in an \astd, then it must be defined by means of \eventB substitutions). 

An interesting refinement is illustrated by the second refinement step. It corresponds to a joint 
refinement of both parts of the model. Not only several actions are put together to form a new one 
(a kind of \emph{merge}) which is consistent in terms of postcondition, but the main behaviour is 
modified such that this new action becomes a synchronisation barrier for the different trains, 
thus implying a behavioural refinement.  

We now aim at generalising all these lessons learnt in order to define a complete methodology, that generates
a readable and more easily provable specification. The development of supporting tools is work in progress.
 
\bibliographystyle{splncs}
\bibliography{Biblio}

\end{document}